\documentstyle[epsfig]{mn}
\ifCUPmtlplainloaded \else

\fi
\title[\textit{HST} Observations of GRO~J0422+32]
      {\textit{Hubble Space Telescope} Observations of the 
       Black Hole X-ray Transient GRO~J0422+32 Near Quiescence}
\author[Hynes \& Haswell]
       {R.I.~Hynes, C.A.~Haswell\thanks{\protect\raggedright Previous address: 
        Columbia Astrophysics Laboratory, Columbia University, 
        538 West 120th Street, New York, NY 10027, USA.} \\
Astronomy Centre, University of Sussex, Falmer, Brighton BN1 9QJ}
\begin{document}
%
%
\newcommand{\novacyg}{V404~Cyg}
\newcommand{\novasco}{GRO\,J1655--40}
\newcommand{\novaper}{GRO\,J0422+32}
\newcommand{\novamus}{X-ray Nova Muscae 1991}
\newcommand{\novamon}{A0620--00}
%
%
\newcommand{\HST} {\textit{HST}}
\newcommand{\XTE} {\textit{RXTE}}
\newcommand{\RXTE}{\textit{RXTE}}
\newcommand{\GRO} {\textit{GRO}}
\newcommand{\ASCA}{\textit{ASCA}}
%
%
\newcommand{\HI}   {H\,\textsc{i}}
\newcommand{\HII}  {H\,\textsc{ii}}
\newcommand{\HeI}  {He\,\textsc{i}}
\newcommand{\HeII} {He\,\textsc{ii}}
\newcommand{\HeIII}{He\,\textsc{iii}}
\newcommand{\MgII} {Mg\,\textsc{ii}}
%
%
\newcommand{\EBV}{E(B-V)}
\newcommand{\Rv} {R_{\rm V}}
\newcommand{\Av} {A_{\rm V}}
%
%
\newcommand{\SXT}{BHXRT}
%
%
\newcommand{\lam}   {$\lambda$}
\newcommand{\lamlam}{$\lambda\lambda$}
\maketitle
%
%
\newcommand{\comm}[1]{\textit{[#1]}}
%
%
\begin{abstract}
We present \HST/FOS ultraviolet and optical spectroscopy of the black
hole X-ray transient \novaper\ shortly before the system reached
quiescence.  We find that the accretion spectrum from 2500--9000\,\AA\
can be very well fit by a self-absorbed synchrotron model, with
superposed \HI\ and \MgII\ emission lines.  The explanations we
suggest for this spectrum are that it is either due to active coronal
regions above a geometrically thin accretion disc, or that the disc is
evaporated into an advective flow.
\end{abstract}
%
%
%
%
\section{Introduction}
\label{intro}
The hard X-ray transient \novaper\ was discovered by \GRO/BATSE on
1992 Aug 5 \cite{P92} at a flux of 0.2 Crab.  It rapidly rose to a
level of 3 Crab, before beginning a gradual decline over subsequent
months.  The X-ray source was identified with a 13th magnitude optical
nova \cite{Castro93}, which showed a similar decline.  The later
stages of the outburst were characterised by an unprecedented series
of mini-outbursts.  Observations in the early stages of its discovery
suggested similarities to other black hole X-ray transient (BHXRT)
binaries; subsequent quiescent studies (Casares et al.\ 1995,
Filippenko et al.\ 1995, Orosz \& Bailyn 1995) have confirmed its
binary nature and suggested a compact object mass of
2.5--5.0\,M$_{\odot}$, larger than any known neutron star and probably
a black hole.  A more recent study has suggested that the mass may
actually be much greater than this \cite{B97}.  In some ways,
\novaper\ is the most extreme member of its class, and the most
similar to cataclysmic variables, having the shortest orbital period
(5.1-hr, Chevalier \& Ilovaisky 1995) and the latest type companion
star (M2V, Casares et al.\ 1995, Filippenko et al.\ 1995).

As part of an ongoing Target-of-Opportunity programme to observe \SXT
s in outburst with The {\it Hubble Space Telescope}, we triggered our
proposal to observe the first of the mini-outbursts.  The observations
were delayed and actually took place two years after the initial
outburst.  The system was nonetheless still active at this point,
being two magnitudes above quiescence in the V band, and we believe we
caught the end of either an otherwise unreported mini-outburst, or an
extended plateau or standstill.

In Sects.~\ref{obs} \& \ref{datred} we describe the observations we
made; they are placed within the context of the long-term light curve
in Sect.~\ref{LongLCSect} and analysed in Sects.~\ref{ContSpecSect} \&
\ref{LineSpecSect}.  Our conclusions are summarised in
Sect.~\ref{conc}.
%
%
\section{Observations}
\label{obs}
\label{hubobs}
\HST\ observed \novaper\ using the Faint Object Spectrograph (FOS) for
two visits, comprising a total of 11 spacecraft orbits, on 1994~August
25 and 26.  The exposures we took are listed in Table~\ref{obstab}.
To observe the vacuum UV, we invested the majority of our time in a
series of G160L exposures, interleaving these with short PRISM
exposures to capture the optical spectrum.  The G270H and G400H
spectra provide coverage overlapping that of the PRISMs at higher
spectral resolution.  All the observations used the `4.3' FOS aperture
and COSTAR.  They were obtained in RAPID mode, with a duty cycle of 77
per cent, and used standard sub-stepping, giving four pixels per
diode.
\begin{table*}
\caption{Journal of \HST\ Faint Object Spectrograph observations of
         \novaper.  The blank line indicates the break between the two
         visits.  The exposure times given exclude dead time, hence
         include only the 77 per cent of the time elapsed.}
\label{obstab}
\begin{tabular}{lcclrc}
                                & 
\multicolumn{1}{c}{Date}        & 
\multicolumn{1}{c}{UT Start}    & 
\multicolumn{1}{c}{Disperser /} & 
\multicolumn{1}{c}{Integration} &
\multicolumn{1}{c}{Wavelength}  \\
\multicolumn{1}{c}{ID}          & 
\multicolumn{1}{c}{1994}        & 
\multicolumn{1}{c}{(hh:mm)}     & 
\multicolumn{1}{c}{Detector}    & 
\multicolumn{1}{c}{Time (s)}    &
\multicolumn{1}{c}{Range (\AA)}  \\
\hline
0104 & Aug 25 & 21:21 & PRISM/RD &    365.4 & 1850--8950\,\AA \\
0105 & Aug 25 & 21:34 & G270H/RD &   1471.1 & 2222--3277\,\AA \\
0106 & Aug 25 & 22:56 & PRISM/RD &    365.4 & 1850--8950\,\AA \\
0107 & Aug 25 & 23:09 & G400H/RD &   1551.8 & 3235--4781\,\AA \\
0108 & Aug 26 & 00:38 & PRISM/BL &    365.4 & 1500--6000\,\AA \\
0109 & Aug 26 & 00:51 & G160L/BL &   1124.7 & 1140--2508\,\AA \\
010a & Aug 26 & 02:10 & G160L/BL &   2273.2 & 1140--2508\,\AA \\
010b & Aug 26 & 04:08 & PRISM/BL &    365.4 & 1500--6000\,\AA \\
\\
0204 & Aug 26 & 15:24 & PRISM/BL &    365.4 & 1500--6000\,\AA \\
0205 & Aug 26 & 15:38 & G160L/BL &    365.4 & 1140--2508\,\AA \\
0206 & Aug 26 & 16:37 & G160L/BL &   2273.2 & 1140--2508\,\AA \\
0207 & Aug 26 & 18:14 & G160L/BL &   1120.0 & 1140--2508\,\AA \\
0208 & Aug 26 & 18:43 & PRISM/BL &    365.4 & 1500--6000\,\AA \\
0209 & Aug 26 & 18:56 & G160L/BL &    142.4 & 1140--2508\,\AA \\
020a & Aug 26 & 19:51 & G160L/BL &   2273.2 & 1140--2508\,\AA \\
020b & Aug 26 & 21:27 & G160L/BL &   1480.6 & 1140--2508\,\AA \\
020c & Aug 26 & 22:01 & PRISM/BL &    365.4 & 1500--6000\,\AA \\
\hline
\end{tabular}
\end{table*}
%
%
\section{Data Reduction}
\label{datred}
Our \HST\ observations found \novaper\ in the faintest post-outburst
state then observed spectroscopically. A preliminary appraisal of the
original pipeline-calibrated data revealed an extremely noisy
spectrum, with only the 2800\,\AA\ \MgII\ emission line detected in
the UV spectrum.  Since the De-commissioning of the FOS, a
substantially revised set of calibration files has become available
\cite{K97}, and we therefore performed a re-reduction and calibration
of our data.  The pipeline reduction is also known to underestimate
the background counts significantly \cite{K95}, necessitating manual
subtraction.
\subsection{G160L Recalibration}
\label{g160lrecal}
\begin{figure}
\begin{center}
\epsfig{width=2in,height=3in,angle=90,file=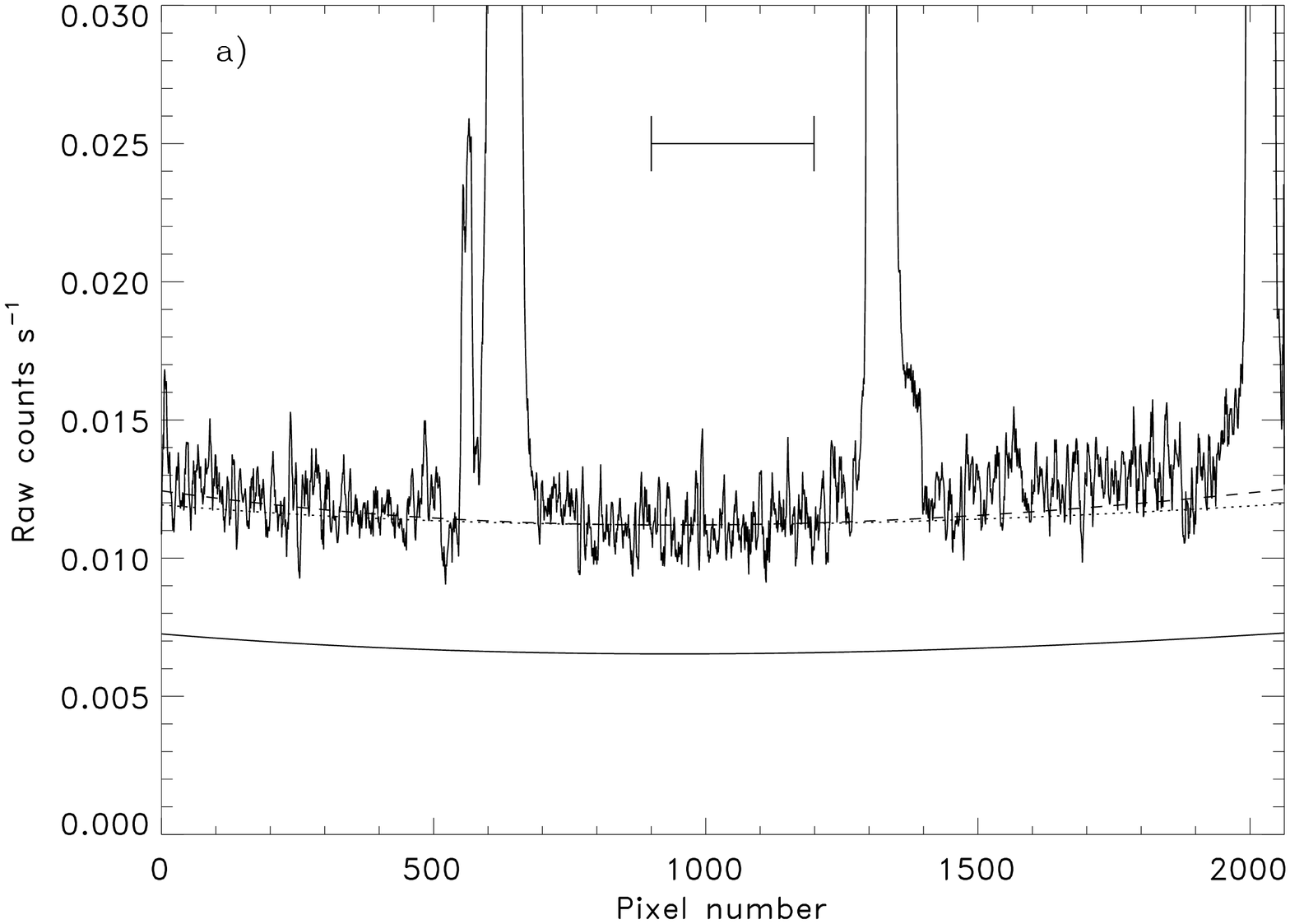}
\epsfig{width=2in,height=3in,angle=90,file=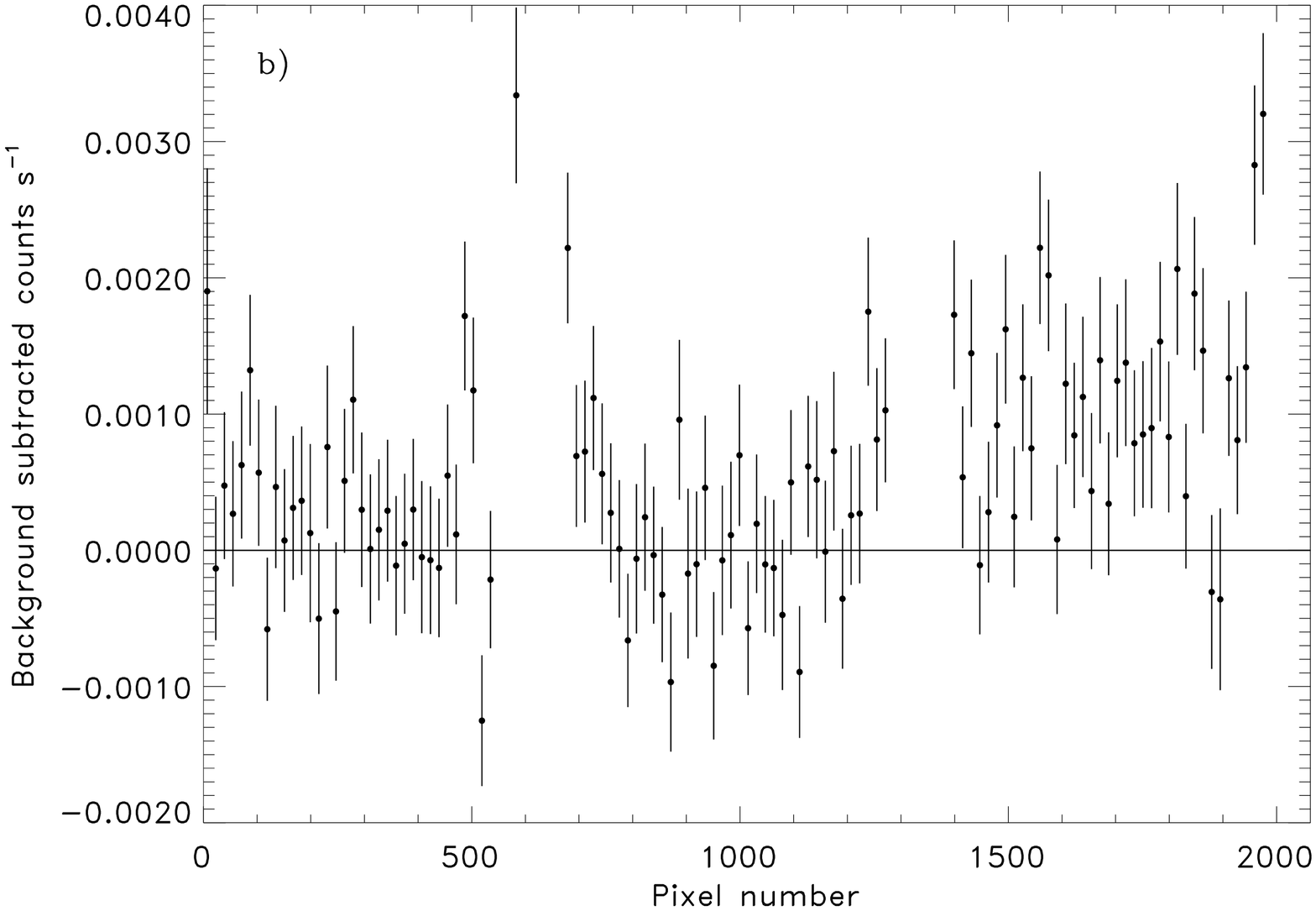}
\epsfig{width=2in,height=3in,angle=90,file=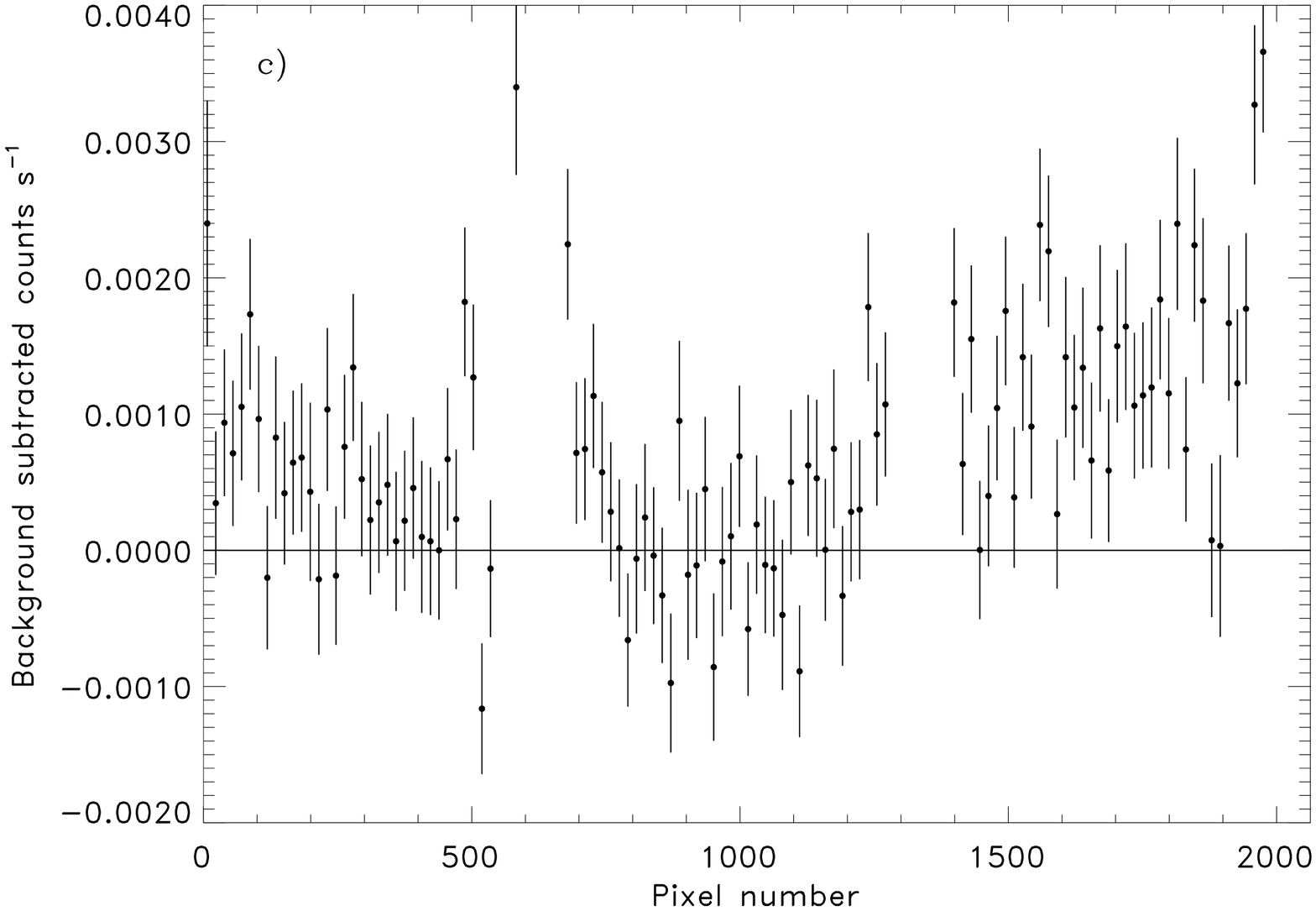}
\caption{a) Raw count rates in our averaged G160L spectrum, over the
         whole detector, smoothed with a 5 pixel boxcar.  We see the
         zeroth order spectrum centred at pixel 640.  The first order
         spectrum runs from pixels 1275--2063.  We clearly see the
         geocoronal Ly$\alpha$ emission around pixel 1320.  We also
         show with the smooth solid line the model background
         calculated based upon the geomagnetic position, and with the
         dashed lines the same scaled up to match the unexposed
         regions (MHR95).  The dotted line shows the model background
         with a fixed (wavelength independent) constant added (KB93).
         The adopted unexposed regions is indicated. b) Rebinned
         ($\times16$) count rates after subtraction using the MHR95
         recipe. c) Rebinned count rates after subtraction using the
         KB93 recipe.}
\label{G160LFig}
\end{center}
\end{figure}

Our average G160L spectrum is shown in Fig.~\ref{G160LFig}.  This
grating disperses the 1140--2508\,\AA\ spectrum to pixels 1275--2063.
Zeroth order light falls on the detector at around pixels 600--670.
Elsewhere in the array, the pixels monitor the background count-rate.
A cursory inspection of our data revealed that the pipeline reductions
had underestimated the background level by approximately a factor of
two.

McClintock, Horne \& Remillard (1995, hereafter MHR95) and Kinney \&
Bohlin (1993, hereafter KB93) disagree about the source of the extra
background light.  MHR95 adopted the hypothesis that it is due to
cosmic rays, whereas KB93 assert that it is due to scattered light.
This disagreement is significant because the {\it shape} of the extra
background spectrum has an effect on the UV continuum flux attributed
to the target.  We believe that in our case the background is
dominated by particle events, in line with the MHR95 interpretation,
for the following reasons:
\begin{enumerate}
\item
MHR95 explicitly ruled out scattered {\em optical} light in their
G160L observations of \novamon.  At the time of our observations,
\novaper\ was two visual magnitudes fainter than the quiescent
magnitude of \novamon, thus scattered optical light from the source is
surely also negligible in our case.
\item
There is very little UV flux from the source so scattering of this is
utterly implausible.
\item
The correction needed for the PRISM/BL spectrum, which used the same
detector, but a different disperser, was of comparable size to that
for G160L.  This suggests that the background level is a property of
the detector, independent of which disperser is used and so supports a
model based on particle events rather than scattered light.  This is
in contrast to the \HST/FOS observations of the highly reddened \SXT\
\novasco\ \cite{Hy98} which probably were influenced by scattered red
light leading to G160L corrections significantly greater than for the
other dispersers.
\end{enumerate}

The one remaining possible source of background counts is scattered
geocoronal light.  Since we used a larger aperture than did MHR95,
this cannot readily be ruled out, but argument (iii) suggests it does
not dominate.

In our grand sum G160L exposure, the background level, determined from
pixels 900--1199, was $71\pm2$ per cent higher than the pipeline
value, as shown in Fig.~\ref{G160LFig}.  The panels below show the
rebinned spectrum obtained by subtracting the revised backgrounds:
Fig.~\ref{G160LFig}(b) uses the MHR95 background recipe,
Fig.~\ref{G160LFig}(c) uses KB93's recipe.  At the left of these
figures can be seen the background subtracted count-rate for the first
546 pixels in the array. These pixels are exposed to background only,
but were not used in the calculation of the background level.
Therefore they provide an assessment of the accuracy of the revised
background.  For both recipes we see significantly lower counts in
this region than from the source region at the right, (pixels
1400--2000) therefore we cautiously conclude that we have detected
vacuum UV emission from \novaper\ in early quiescence.  The revised
background subtraction shown in panel (b) gives a mean count rate per
pixel in pixels 0--545 of $(2.8\pm0.9)\times10^{-4}$\,s$^{-1}$, about
half that in panel (c), $(5.4\pm0.9)\times10^{-4}$\,s$^{-1}$.  While
the particle model seems to give a better estimate of the shape of the
background count rate than a scattered light model, neither model
gives a mean count in pixels 0--545 within $3\sigma$ of zero so we
should view these results with considerable caution, as this may
indicate a deficiency in {\em both} models.
\subsection{PRISM Recalibration}
\label{PrismRecalSect}
As for G160L, for both PRISM configurations (BL and RD) we combined
exposures into a grand sum before attempting to characterise the
background.  Both PRISM configurations give many pixels exposed to
background only: we used pixels 1400--1999 for PRISM/BL and 100--899
for PRISM/RD.  For the former case, the observed background count rate
was higher than the background model by $63\pm3$ per cent, a similar
value to that found with G160L which also used the BL detector.  For
PRISM/RD no significant correction was needed; the background model
background was $2\pm3$ per cent below that observed.

For the PRISM/RD combination, there appeared to be an offset in the
dispersion direction, probably due to miscentering of the target in the
aperture.  We detected this by comparison of the position of the
\MgII\ (2798\,\AA) line with its position in the higher resolution
G270H spectrum.  The offset was the same for both PRISM/RD exposures,
so the miscentering was probably also the same for the G270H exposure
taken between them.  We were able to correct the offset by shifting
both G270H and PRISM/RD spectra by 12 pixels after flat-fielding.
This also brought the H$\alpha$ and H$\beta$ emission lines in the
PRISM/RD spectrum to within one pixel of their rest wavelengths.  Since
the PRISM/BL spectrum contains no prominent emission lines (the \MgII\
line is undetected), we cannot determine if there is a similar offset
in this case.  This makes our calibration of the PRISM/BL spectra less
certain than PRISM/RD.  Since the BL detector is also less sensitive
than the RD, we focussed on the PRISM/RD spectrum in our model
fitting.

\begin{figure}
\begin{center}
\epsfig{width=2in,height=3in,angle=90,file=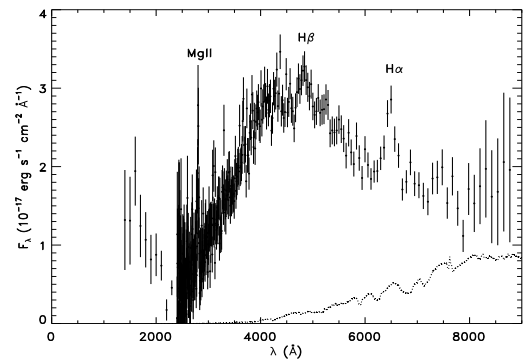}
\caption{G160L and PRISM/RD spectra.  The G160L spectrum has been
         averaged into 100\,\AA\ bins.  Also shown dotted is our
         estimate of the spectrum of the companion star (see
         Sect.~\ref{SecSect}), reddened by $\EBV=0.3$.}
\label{CompSpecFig}
\end{center}
\end{figure}

We note that a 12 pixel error would correspond to 4 diodes or 1.22
arcsec in the mode used.  Since the aperture width was 4.3 arcsec, the
stellar image should still have been $\sim1$ arcsec, or $10\sigma$,
from the aperture edge and so aperture losses will not significantly
affect the accuracy of the spectrophotometry.  We also note that the
revised calibration files \cite{K97} made a significant difference for
the extreme long wavelength end of the PRISM/RD spectrum, with the
flux longwards of 7500\,\AA\ being increased by 20--80 per cent.  Our
recalibrated G160L and PRISM/RD spectra, spanning 1300--9000\,\AA\ are
shown in Fig.~\ref{CompSpecFig}.
\subsection{G270H and G400H Recalibration}
\label{G270HRecalSect}
Because of the high dispersion, the count rates are relatively low and
the background is a serious problem for both G270H and G400H.  The
problem is exacerbated because these gratings both completely fill the
diode array with the source spectrum, leaving no region to monitor the
background.  Comparing these spectra with those from the PRISMs, the
pipeline does seem to have subtracted the background reasonably well,
but we cannot attempt a precise recalibration of the high resolution
spectra or quantify the uncertainty in the background level.  We
therefore use the G270H and G400H spectra primarily to search for and
examine spectral lines.

\begin{figure}
\begin{center}
\epsfig{width=2in,height=3in,angle=90,file=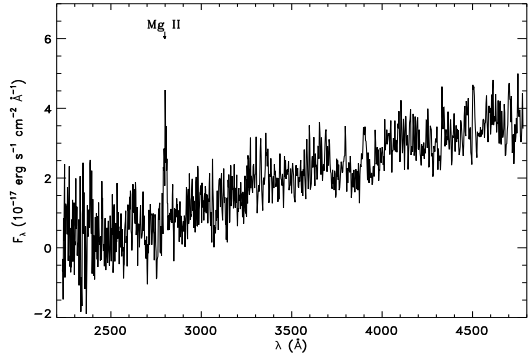}
\caption{Combined G270H and G400H spectrum.  These have been convolved
         with a Gaussian of width equal to the line spread function
         and rebinned for clarity.  The spectrum is largely
         featureless, with the \MgII\ (2798\,\AA) emission line the
         only prominent feature.}
\label{HiResFig}
\end{center}
\end{figure}

Our combined G270H and G400H spectrum is shown in Fig.~\ref{HiResFig}.
For clarity it has been convolved with a Gaussian of the same width as
the estimated line spread function, i.e.\ one diode.  This corresponds
to 2.05\,\AA\ for the G270H grating and 3.00\,\AA\ for the G400H.  The
smoothed spectrum was then rebinned to give one pixel per resolution
element.  The spectrum is largely featureless with only the \MgII\
(2798\,\AA) emission line unambiguously detected.
%
%
\section{The Photometric State of \novaper}
\label{LongLCSect}
\begin{figure}
\begin{center}
\epsfig{width=2in,height=3in,angle=90,file=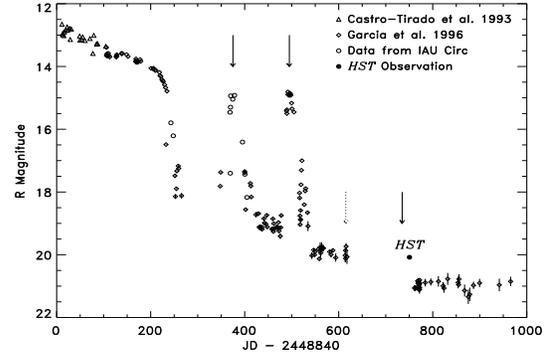}
\caption{The R band lightcurve of the outburst, adapted from Fig.\ 1
         of Garcia et al.\ (1996).  Vertical arrows indicate observed
         or extrapolated times of mini-outburst.  Our measurement
         clearly lies significantly above the subsequent photometry,
         and is consistent with the expected time of a mini-outburst.}
\label{PhotFig}
\end{center}
\end{figure}

The R band light curve of the outburst of \novaper\ is shown in
Fig.~\ref{PhotFig}.  On 1994 September 5, approximately 10 days after
our observations, Zhao et al.\ \shortcite{Z94b} measured magnitudes of
$V=22.39\pm0.27$, $R=21.06\pm0.10$.  The system remained at this level
and this is representative of its subsequent quiescent state
($V=22.24\pm0.14$, $R=20.97\pm0.10$, Casares et al.\ 1995.)  We can
also calculate approximate magnitudes from our PRISM/RD spectrum
however, and obtain $U=20.75\pm0.04$, $B=20.84\pm0.03$,
$V=20.42\pm0.04$, $R=20.08\pm0.04$, brighter by one magnitude in the R
band and two magnitudes in V.  This is a significant difference, and
it is very unlikely that the \HST\ calibration could be this much in
error; our error estimates include statistical errors, the documented
3--4 per cent photometric accuracy of the FOS \cite{K97} and allow for
a miscentering uncertainty $\pm1$ pixel.  We conclude that \novaper\
was not completely quiescent at the time of our observations.
Instead, we suggest that we saw the final stages of a last, previously
undetected, mini-outburst.  If this is the case, then this
mini-outburst occurred approximately 240 days after the last recorded
one, consistent with the recurrence time of 120 days originally
suggested for the 1993--94 mini-outbursts (Augusteijn, Kuulkers \&
Shaham 1993, Chevalier \& Ilovaisky 1995) and subsequently detected in
quiescence (Iyudin \& Haberl 1997a,b).  An alternative interpretation
is that after the last observations around day 600, when Zhao et
al.\ \shortcite{Z94a} measured $R=20.03\pm0.11$, $V=20.67\pm0.22$, the
system remained on a plateau, finally dropping to quiescence just
after our observation.  This is plausible, as both R and V band
magnitudes around day 600 are similar to values we observe.
%
%
\section{The Continuum Spectrum}
\label{ContSpecSect}
\subsection{Interstellar Reddening}
Due to the low signal-to-noise ratio and uncertain background level of
our G160L spectrum, we cannot reliably attempt our own reddening determination
using the 2175\,\AA\ feature.  Existing reddening estimates are
summarised in Table \ref{ReddenTable}.  We adopt a compromise value of
$\EBV=0.3\pm0.1$ and deredden our spectra using the galactic average
extinction curve of Seaton \shortcite{S79}.  Since the only UV
determination lies at the extreme end of the spread of estimates,
however, it may be that the extinction curve actually diverges from
the galactic average; our dereddened UV spectrum should therefore be
viewed with some caution.
\begin{table}
\caption{Estimates of the reddening of \novaper.  References are 1.\
         Shrader et al.\ \shortcite{S94}, 2.\ Callanan et al.\
         \shortcite{C95}, 3.\ Harlaftis \& Charles \shortcite{HC93},
         4.\ Chevalier \& Ilovaisky \shortcite{CI95}, 5. Castro-Tirado
         et al.\ \shortcite{Castro93}}
\label{ReddenTable}
\begin{tabular}{llc}
Method                   & Value         & Ref. \\
\hline
2175\,\AA\ feature       & $0.40\pm0.06$ & 1.\\
EW of interstellar lines & $0.2\pm0.1$   & 1.\\
                         & $0.3\pm0.1$   & 2.\\
                         & $0.2$         & 3.\\
Surrounding field stars  & $0.40\pm0.07$ & 4.\\
Optical continuum shape  & $0.23\pm0.02$ & 1.\\
Unstated                 & $0.3\pm0.1$   & 5.\\
\hline
\end{tabular}
\end{table}
\subsection{The Companion Star Contribution}
\label{SecSect}
Since \novaper\ was near quiescence when we observed it, the spectrum
of the companion star can be expected to be a significant fraction of
the total, at least at
longer wavelengths.  Unfortunately, the resolution of the PRISM/RD
spectrum at long wavelengths is too low to detect spectral features
from the companion, so we cannot perform an independent analysis of
veiling to estimate how strong the contribution is.  Instead we adopt
results from the literature.

\begin{figure}
\begin{center}
\epsfig{width=2in,height=3in,angle=90,file=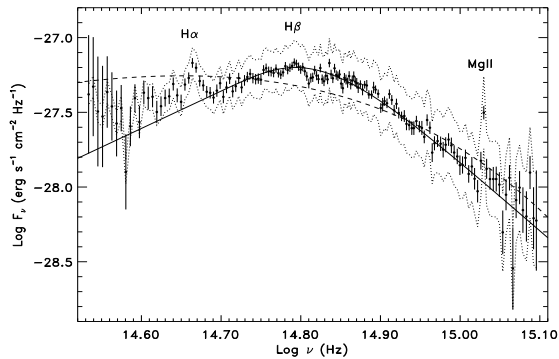}
\caption{Dereddened PRISM/RD spectrum with the companion contribution
         subtracted.  Points with error bars correspond to the
         $\EBV=0.3$ solution; the effect of instead assuming
         $\EBV=0.2$ or 0.4 is indicated by dotted lines.  The data
         below 4000\,\AA\ ($\log\nu>14.88$) has been averaged into
         25\,\AA\ bins for clarity.  The large error bars at low
         frequencies are systematic and arise from assuming a $\pm1$
         pixel uncertainty in the miscentering correction.  The best
         fitting black body (dashed) and self-absorbed synchrotron
         (solid) spectra are overplotted.}
\label{SynchFitFig}
\end{center}
\end{figure}

For the spectral type of the companion star we assume M2V as favoured
by Casares et al.\ \shortcite{C95} and Filippenko et al.\
\shortcite{F95} and use the spectrum of the M2V star GL49 \cite{GS83}
as the template for our companion contribution.  Casares et al.\
\shortcite{C95} give an R band (quiescent) magnitude of $20.96\pm0.10$
and estimate the fraction of the light due to the secondary to be
$52\pm8$ per cent in the 6700--7500\,\AA\ range.  Their results are
consistent with those of Filippenko et al.\ \shortcite{F95}.  We
therefore estimate an R band magnitude for the companion star alone of
$21.68\pm0.11$ and normalise the spectrum of GL49 accordingly; the
resulting spectrum, reddened by $\EBV=0.3$, is shown in
Fig.~\ref{CompSpecFig}.  We then subtract this normalised spectrum,
smoothed to the instrumental resolution of $1\,{\rm diode}=4\,{\rm
pixels}$ \cite{K97}, to obtain the spectrum of the accretion flow
shown in Fig.~\ref{SynchFitFig}.

\novaper\ presents a unique opportunity to study the spectrum of the
quiescent or near quiescent accretion flow in a \SXT.  Because the
companion star is an M-type dwarf, rather than the more typical K, G
or even F-type subgiant, it is redder than in the other systems and
contaminates the optical spectrum much less.  Furthermore the late
type companion shows very strong molecular bands, so its spectral type
and brightness can be better determined, and so the uncertainty in the
contamination is smaller.  The deduced spectrum of the accretion flow
is therefore much less affected by uncertainties in the companion's
contribution.
\subsection{Black Body Models}
We began by trying to characterise the spectrum of the accretion flow
with a black body fit.  The best fitting black body spectrum (in the
$\chi^2$ sense) is shown by the dashed line in Fig.~\ref{SynchFitFig}.
This has $T \sim 7600$\,K and a reduced $\chi^2$ of 3.6 after masking
out emission lines.  It is clear that this is not a satisfactory fit
and allowing for a $\pm0.1$ uncertainty in $\EBV$ does not resolve
this.  The observed spectrum is sharper peaked than a black body and
tends to a power-law form at short wavelengths.  Adopting a
temperature distribution in a disc instead of a single temperature
black body can produce a power-law at short wavelengths, but will
always produce a broader spectrum than a single temperature model.  In
principle, a thermal spectrum from an accretion disc might be expected
to deviate from a black body, but we would expect deviations in the
form of edges or structure into the spectrum, in particular a Balmer
jump in emission or absorption.  Apart from the sharp, and uncertain,
upturn below 2500\,\AA\ (Fig.~\ref{CompSpecFig}), there is no evidence
for this.  We encountered similar problems in trying to fit the
spectrum of \novasco\ in outburst, which also seemed too sharply
peaked to be fit by a black body model \cite{Hy98}.  There, however,
the situation was complicated by the much larger reddening.  In this
case we have much less uncertainty in isolating the intrinsic spectrum
of the accretion flow, and hence we can say with more confidence that
it cannot be modelled by purely thermal emission.
\subsection{Non-thermal Models}
\label{AdvectionSect}
A more promising explanation for the observed spectrum is that it is
non-thermal in origin: self-absorbed synchrotron emission.  Such a
mechanism can readily produce a spectrum with a sharper peak than a
black body and, as illustrated in Fig.~\ref{SynchFitFig}, can fit the
spectrum rather well.  The spectrum plotted is based on a free-fit of
a synchrotron emitting slab containing a magnetic field of uniform
strength and electrons with a power-law energy spectrum.  This fit is
intended purely to demonstrate that even a very simple synchrotron
emitting model can fit the accretion spectrum well.  More realistic
models of self-absorbed synchrotron emission in X-ray binaries invoke
thermal electron distributions, but over the limited spectral range
covered, a power-law model leads to a reasonable approximation to the
expected spectrum.

We suggest two more sophisticated models that could lead to such a
spectrum.  The first is that the accretion disc extends inwards close
to the compact object but has a magnetically dominated corona above
it.  Such a model was considered by Di Matteo, Celotti \& Fabian
\shortcite{DMCF97} in the context of both active galactic nuclei and
galactic black hole candidates.  In their model there are at any time
a number of localised active coronal regions in which electrons are
energised by magnetic reconnection (a possible mechanism for this is
described by Haswell, Tajima \& Sakai 1992) and subsequently, cool
emitting cyclo-synchrotron radiation which becomes self-absorbed below
some critical frequency.  These models predict that for solar mass
black holes, the critical frequency, and hence the peak of the
spectrum, should lie in the extreme UV, i.e.\ at higher frequencies
than we observe.  We are observing \novaper\ in a much lower state
than was considered by Di Matteo et al.\ \shortcite{DMCF97}, however,
and this could move the peak to lower energies.

An alternative set of models, which have seen much discussion
recently, are the so called advective accretion flows.  These have
been put forward as an explanation of many properties of \SXT s in
quiescence (Narayan, McClintock \& Yi 1996, Hameury et al.\ 1997,
Narayan, Barret \& McClintock 1997), and subsequently the models have
been extended to consider the outburst and decline phases
\cite{EMN97}.  The essence of these models is that the inner region of
the disc becomes evaporated into a geometrically thick, optically thin
region.  This flow radiates less efficiently than an optically thick disc
and so there is a bulk inflow of thermal energy (i.e.\ the flow is
advective), ultimately through the event horizon.  The models predict
that the optical and UV accretion light should be dominated by
self-absorbed synchrotron emission from this advective inner region.  Since our
spectrum was obtained at a very low luminosity, shortly before
\novaper\ reached its quiescent state, we should consider if these
models agree with our observations.  A detailed comparison is beyond
the scope of this paper, requiring some refinement of the models, but
a preliminary analysis appears promising (A. Esin, priv.\ comm.).  Both
the frequency of the predicted synchrotron peak and its luminosity are
strikingly consistent with our observations.  We therefore suggest it
is possible that the optical continuum emission at the time of our
observations was dominated by an advective flow with mass transfer
rate somewhat above quiescence.
\subsection{The Far Ultraviolet Spectrum}
As discussed in Sect.~\ref{g160lrecal}, we appear to have detected
source counts in the far UV from \novaper.  For a range of extinctions
$\EBV=0.2-0.42$, our average flux over the range 1650--1950\,\AA, is
in the range $f_{\lambda} = (0.4-1.9) \times
10^{-16}$\,erg\,s$^{-1}$\,cm$^{-2}$\,\AA$^{-1}$.  Normalised to what
would be seen at a distance of 1\,kpc (assuming a distance of 2.4\,kpc
for \novaper, Shrader et al.\ 1994), $f_{\lambda} = (3-11)
\times10^{-16}$\,erg\,s$^{-1}$\,cm$^{-2}$\,\AA$^{-1}$.  Uncertainties
both in the {\em shape} of the background spectrum and its
normalisation will increase this range and our results do not
conclusively rule out zero flux in the far UV.  Comparing this with
the normalised far UV fluxes tabulated by MHR95 for \novamon\ and
selected dwarf novae, our measurement agrees with the latter, but is
ten times greater than in \novamon.  This may be because \novaper\ was
not completely quiescent at the time of our observation, but since our
detection is marginal, the disagreement may not be significant.  Our
deduced far UV spectrum is strongly dependent on the exact form and
level of background spectrum adopted.  The data indicate a sharp
upturn in the UV, however the reliability of this is far from certain
(Fig.\ \ref{G160LFig}).  If it is indeed a real upturn, it would not
readily come from advective models in their current form (e.g.\ Esin
et al.\ 1997).  While these do predict an upturn due to Comptonisation
of the synchrotron peak, this is expected to occur at shorter
wavelengths and to be smoother than we seem to observe.
%
%
\section{Emission Lines}
\label{LineSpecSect}
\begin{table}
\caption{Dereddened fluxes and full width half maxima of detected
         emission lines.  The measured width of H$\alpha$
         is comparable to the width of the line spread function (LSF),
         so is not meaningful.  The fit to H$\beta$ was very poor.}
\label{LineTable}
\begin{tabular}{lccc}
 & Line flux                  & Line & \multicolumn{1}{c}{LSF} \\
 & (erg\,s$^{-1}$\,cm$^{-2}$) & FWHM (\AA) & 
\multicolumn{1}{c}{FWHM (\AA)} \\
\hline
\MgII (2798) & $2.8 \pm 0.4\times10^{-15}$ & $18.5\pm2.4$ & 2 \\
H$\beta$          & $3.4 \pm 0.6\times10^{-15}$ & $240\pm50$   & 120\\
H$\alpha$         & $3.8 \pm 0.7\times10^{-15}$ & ($200\pm30$) & 270\\
\hline
\end{tabular}
\end{table}
In the optical, we detect unresolved H$\alpha$ and barely resolved
H$\beta$ emission in our PRISM/RD spectrum.  The only emission line
detected in the ultraviolet was \MgII\ (2798\,\AA), which was also
seen to be prominent in \novamon\ (MHR95).  The deduced line
parameters are given in Table \ref{LineTable}.  The \MgII\ line is
well resolved by the G270H grating and has a FWHM in velocity space of
$2000\pm300$\,km\,s$^{-1}$, comparable to that observed in \novamon.
We estimate that emission is detectable to velocities of $\sim
1500$\,km\,s$^{-1}$ (i.e.\ ${\rm FWZI}\sim3000$\,km\,s$^{-1}$.  This
is similar to the quiescent H$\alpha$ and H$\beta$ lines observed by
Orosz \& Bailyn \shortcite{OB95}, for which we estimate FWHM
1500\,km\,s$^{-1}$ and 1800\,km\,s$^{-1}$ respectively, with FWZI
$\sim 3000$\,km\,s$^{-1}$ in both cases.

If we assume the system parameters of Filippenko et al.\
\shortcite{F95}, then at the circularisation radius in the disc we
would expect $v_{\phi} \sin i \sim 700$\,km\,s$^{-1}$.  The observed
velocities of the \MgII\ emission (and the quiescent H$\alpha$
emission) therefore suggest emission from a disc extending inwards to
at least one quarter the circularisation radius.  This is consistent
with the advective accretion models discussed in
Sect.~\ref{AdvectionSect}, which assume evaporation of the accretion
disc at a few times less than the circularisation radius \cite{EMN97},
but obviously this does not prove that the disc is evaporated.
%
%
\section{Summary and Conclusions}
\label{conc}
We have presented a study of \novaper\ near quiescence based on UV and
optical spectroscopic observations, and suggest that these took place
during the decline from a previously unreported mini-outburst.  Our
observations show similarities to other \SXT s in {\em quiescence},
and in particular to \novamon, in two respects:
\begin{enumerate}
\item
We find that this accretion spectrum cannot be well fitted by black
body models, but can be better characterised by a self-absorbed
synchrotron spectrum.  Two possible explanations have been advanced:
either a magnetically-dominated corona above a thin disc, or an
advection dominated flow.  Further modelling beyond the scope of this
paper will be required to discriminate between these possibilities.
\item
The emission line spectrum is dominated by H$\alpha$, H$\beta$ and
\MgII\ (2798\,\AA), as is found for \novamon\ in quiescence.  The
\MgII\ width is similar to that in \novamon, suggesting a similar
origin.
\end{enumerate}

The similarity on both counts to other quiescent \SXT s suggests a
similar structure.  The brightness of \novaper\ at the time of our
observations can be readily explained if the mass transfer rate had
not completely dropped to its quiescent level.
%
%
\section*{Acknowledgements}
RIH is supported by a PPARC Research Studentship.  Thanks to our
referee, Tim Naylor, and the editor, Andy Fabian, for constructive
criticism and advice, and to Phil Charles, Erik Kuulkers and Tariq
Shahbaz for hospitality, and helpful suggestions and discussion on
this work.  The advice of Tony Keyes is appreciatively acknowledged,
as is the generously shared expertise of the STSDAS team at STScI.  We
thank Ping Zhao, Paul Callanan, Mike Garcia and Jeff McClintock for
obtaining photometry shortly after our observations and providing
their full set of photometry for inclusion in Fig.~\ref{PhotFig}.  We
used the NASA Astrophysics Data System Abstract Service and support
for this work was provided by NASA through grant number GO-4377-02-92A
from the Space Telescope Science Institute, which is operated by the
Association of Universities for Research in Astronomy, Incorporated,
under NASA contract NAS5-26555.
%
%

%
\end{document}